\documentclass[preprint]{vgtc}               

\graphicspath{{figures/}{pictures/}{images/}{./}} 

\usepackage{times}                     
\usepackage{comment}

\usepackage{tabu}                      
\usepackage{booktabs}                  
\usepackage{lipsum}                    
\usepackage{mwe}                       
\usepackage {algorithm}
\usepackage{amsmath,amssymb,amsfonts}
\usepackage{mathptmx}                  

\usepackage[inline]{enumitem}

\onlineid{0}

\vgtccategory{Posters}

\vgtcinsertpkg

\preprinttext{Accepted as a poster at the IEEE VIS Conference 2025. "TS-Insight: Visual Fingerprinting of Multi-Armed Bandits"}

\title{TS-Insight: Visualizing Thompson Sampling for Verification and XAI}

\author{Parsa Vares\thanks{e-mail:parsa.vares@list.lu}\\ %
        \scriptsize University of Luxembourg\\
        \scriptsize Luxembourg Institute of Science and Technology
\and \'Eloi Durant\thanks{e-mail:eloi.durant@list.lu}\\ %
        \scriptsize Luxembourg Institute of Science and Technology %
\and Jun Pang\thanks{e-mail:jun.pang@uni.lu}\\ %
        \scriptsize University of Luxembourg %
\and Nicolas Médoc\thanks{e-mail:nicolas.medoc@list.lu}\\ %
        \scriptsize Luxembourg Institute of Science and Technology %
\and Mohammad Ghoniem\thanks{e-mail:mohammad.ghoniem@list.lu}\\ %
        \scriptsize Luxembourg Institute of Science and Technology} %

\teaser{
  \centering
  \includegraphics[width=0.90\linewidth]{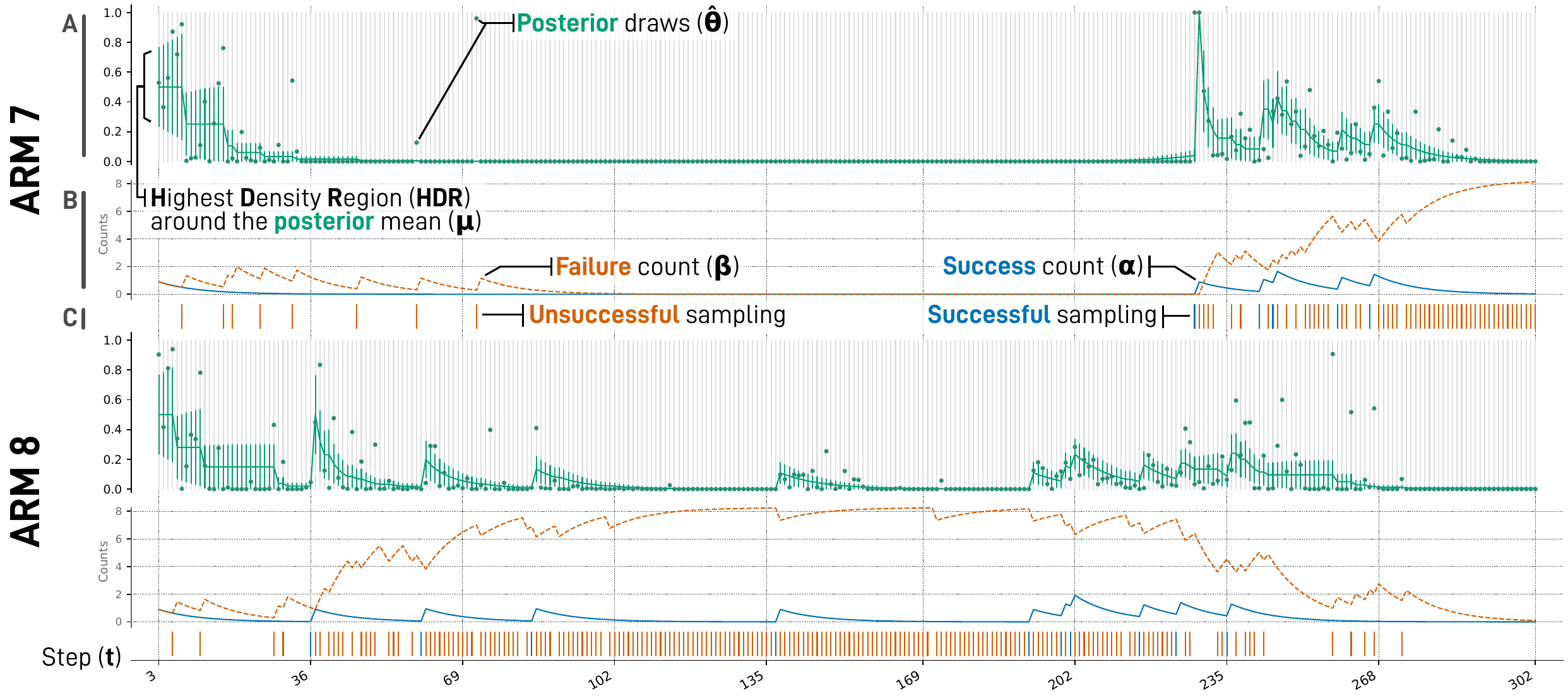} 
    \caption{Excerpt from the TS-Insight dashboard, illustrating the behavior of multiple arms over time, each decomposed in three subplots. \textbf{A}) reveals the algorithm's belief and uncertainty about an arm's success rate. \textbf{B}) shows the raw evidence accumulation. \textbf{C}) provides a compact history of when an arm was chosen and if its sampling succeeded (blue) or failed (orange). These depict, for instance, how an arm's belief narrows as more evidence is gathered. Arms and subplots can be displayed or hidden at will.}
  \label{fig:teaser}
}

\abstract{
Thompson Sampling (TS) and its variants are powerful Multi-Armed Bandit algorithms used to balance exploration and exploitation strategies in active learning. Yet, their probabilistic nature often turns them into a ``black box'', hindering debugging and trust. We introduce TS-Insight, a visual analytics tool explicitly designed to shed light on the internal decision mechanisms of Thompson Sampling-based algorithms, for model developers. It comprises multiple plots, tracing for each arm the evolving posteriors, evidence counts, and sampling outcomes, enabling the verification, diagnosis, and explainability of exploration/exploitation dynamics. This tool aims at fostering trust and facilitating effective debugging and deployment in complex binary decision-making scenarios especially in sensitive domains requiring interpretable decision-making.

} 


\begin{document}

\firstsection{Introduction}
\maketitle

Thompson Sampling (TS)~\cite{Thompson1933, Russo2018} and its variants, e.g., Discounted TS (DTS)~\cite{Raj_DTS_NeurIPS2017}
are state-of-the-art for sequential decision-making under uncertainty. Their many applications include A/B testing, clinical trials and Algorithm Selection for active learning~\cite{zhang2023tailor}. Despite their success, their probabilistic mechanics are often opaque, hindering debugging, tuning, and trust. Existing visualizations typically show aggregate metrics only (e.g., cumulative regret~\cite{Russo2018}), hiding the instance-level dynamics crucial for developers. While platforms like Google Vizier~\cite{GoogleVizier2017} leverage bandit strategies, we still lack visual analytics tools for developers supporting a step-by-step inspection of the internal mechanics of TS-based algorithms.

\begin{figure}[tb]
 \centering 
\includegraphics[width=0.95\columnwidth]{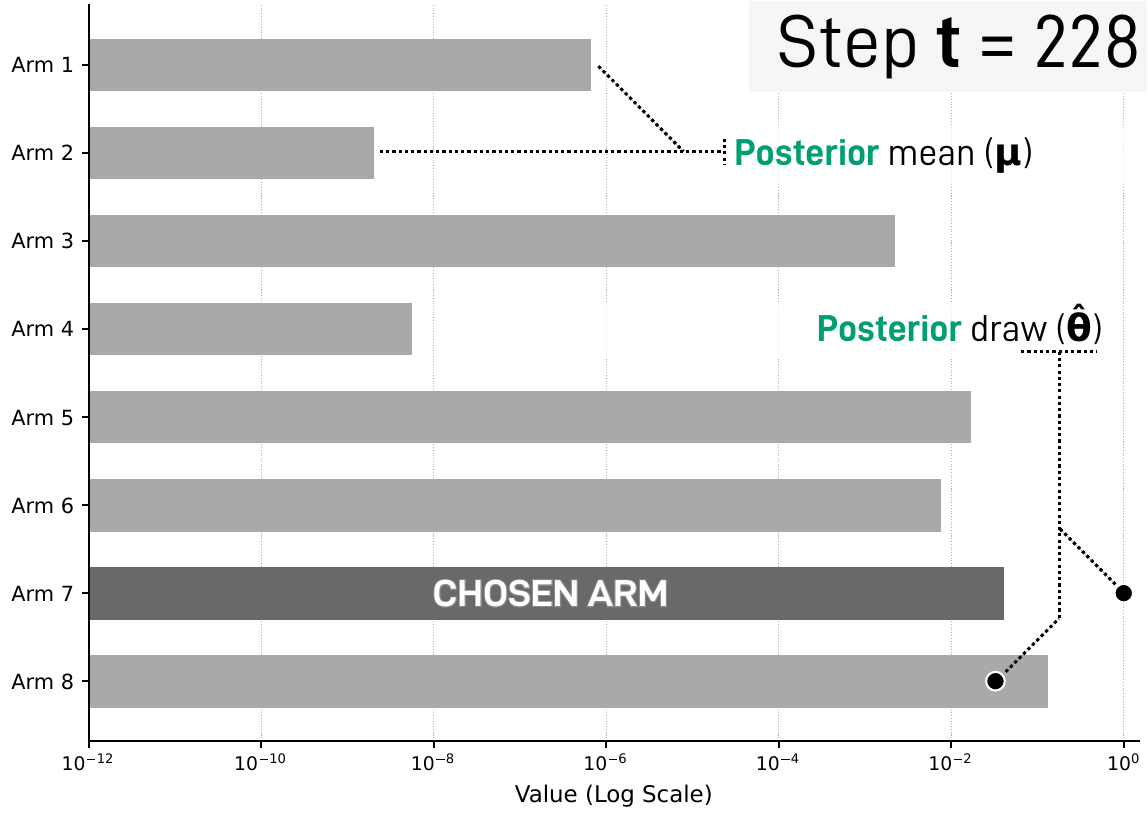} 
 \caption{The XAI Snapshot view explains the algorithm's choice at sampling step ($t=228$). One can see that Arm 8 had the highest \emph{posterior mean}, but Arm 7 was chosen since its \emph{posterior draw} ($\hat{\theta} \approx 1.0$) was the highest, illustrating the \textbf{Exploration} strategy.} \label{fig:xai_snapshot}
\end{figure}

\noindent This paper introduces \textbf{TS-Insight}\footnote{\url{https://github.com/LIST-LUXEMBOURG/ts-insight}}, an open-source web application
built to expose the inner mechanics of TS-based systems. Our contributions are: 
\begin{enumerate*}[label=\textbf{\arabic*.}]
    \item a visual design to verify the mechanics at play within each arm of a Thompson Sampling system (like discounting), composed of three subplots showing A) posterior evolution, B) evidence history, and C) sampling outcomes;
    \item an \textbf{XAI Snapshot view} (\cref{fig:xai_snapshot}) showing the factors involved in any arm selection step; and
    \item a customizable visual dashboard enabling the selective display of chosen arms, related subplots, or sampling step ranges.
\end{enumerate*} 

\noindent Inspired by Zhang~et~al.~\cite{zhang2023tailor}, we use this tool to debug DTS-based active learning applied to the binary classification problem of scientific papers as relevant/irrelevant in the scope of systematic literature review projects. We use the SYNERGY fully labeled paper corpora~\cite{deBruin2023synergy} to run a preliminary validation of TS-Insight.

\section{Background: TS and DTS Mechanics}
\label{sec:Background}
Thompson Sampling (TS)~\cite{Thompson1933, Russo2018} is a Bayesian heuristic for multi-armed bandits problem with binary rewards, modeling each arm’s probability of success $\theta_k$ as a Beta($\alpha_k, \beta_k$) distribution, where $\alpha_k$ and $\beta_k$ are posterior parameters representing pseudo-counts of observed successes and failures, respectively. At each step $t$, it 
\begin{enumerate*}[label=\textbf{\arabic*.}]
    \item draws a sample $\hat{\theta}_{k,t}$ (a \emph{posterior draw}) from each arm's Beta distribution,
    \item selects the arm $a_t = \arg\max_k \hat{\theta}_{k,t}$,
    \item observes a binary reward, and
    \item updates the corresponding $\alpha_{a_t}$ (if success) or $\beta_{a_t}$ (if failure).
\end{enumerate*} 
A \textbf{stationary environment} assumes fixed reward distributions over time~\cite{Russo2018}, i.e., $P(r_t \mid a_t = k)$ remains constant. In a \textbf{non-stationary environment}, these distributions may shift~\cite{Raj_DTS_NeurIPS2017}. \emph{Discounted} TS (DTS) adapts to such changes by applying a discount factor $\gamma \in (0,1]$ to all evidence counts before each update: $\alpha_k \leftarrow \gamma \alpha_k$, $\beta_k \leftarrow \gamma \beta_k$, thus prioritizing recent evidence.
At each step, TS follows one of two strategies:
(1)~\textbf{Exploitation}, when the selected arm has the highest posterior mean ($\mu_k = \alpha_k / (\alpha_k + \beta_k)$),
(2)~\textbf{Exploration}, when a lower-mean arm is selected due to a higher posterior draw $\hat{\theta}_{k,t}$, exploiting variance in the posterior.

\section{TS-Insight: Visualization Design}
\label{sec:design}
To provide a comprehensive understanding of TS behavior, our dashboard presents a composite view for each arm (\cref{fig:teaser}), arranged in rows of up to three synchronized subplots. This separation of concerns supports detailed comparison across arms and time.

\noindent{\textbf{HDR Evolution Plot}\,--\,Belief and Uncertainty:}
To visualize the algorithm’s belief, we use a Highest Density Region (HDR) plot~\cite{hyndman1996computing} (\cref{fig:teaser}A). This plot frames each posterior draw ($\hat{\theta}$)—which drives arm selection—within the belief distribution. The HDR band highlights the most probable region under the current posterior. When a draw falls outside this region, it signals a rare event, revealing decision uncertainty and supporting the analysis of exploratory behavior.

\noindent{\textbf{Alpha/Beta \& Barcode}\,--\,Evidence and History:}
The Alpha/Beta subplot (\cref{fig:teaser}B) tracks cumulative evidence through the posterior parameters $\alpha$ and $\beta$, supporting the verification of belief updates and discounting dynamics in DTS. The Barcode view (\cref{fig:teaser}C) summarizes the arm selection history and outcomes using color-coded strokes. It facilitates the identification of potential non-stationarity in the environment by visualizing shifts in the arms being selected across different sampling phases.

\noindent{\textbf{XAI Snapshot View}\,--\,Explanation:}
For step-level explanation, the XAI Snapshot View (\cref{fig:xai_snapshot}) compares each arm's posterior mean ($\mu$) and posterior draw ($\hat{\theta}$) in a bar chart. This highlights whether a selection followed strong belief or resulted from a high but unlikely draw. The logarithmic scale aids interpretation when dealing with low-probability rewards.

\section{Expected Insights and Explainable AI (XAI)}
\label{sec:insights}

TS-Insight is designed to answer three key questions: \emph{``Is the algorithm working correctly?''} (verification), \emph{``Why did the algorithm make this specific choice?''} (explanation), and \emph{``When was the algorithm’s choice outside the certainty region?''} (reliability).

\noindent{\textbf{Verification of Algorithm Mechanics and Performance:}}
TS-Insight provides direct visual support for algorithm verification. Users can trace a selection in the Barcode view (\cref{fig:teaser}C) to a corresponding change in the Alpha/Beta plot (\cref{fig:teaser}B), confirming the update logic. In DTS, the forgetting mechanism is made visible through the decay of $\alpha$ and $\beta$ for idle arms. The Barcode view also helps assess whether the algorithm adapts to environmental shifts by revealing changes in the distribution of selected arms across sampling phases. For example, at $t=235$ (\cref{fig:teaser}C), Arm~8 successfully selects a relevant item, increasing its $\alpha$ (\cref{fig:teaser}B). No relevant pulls follow, leading to a decay of its $\alpha$ due to discounting.

\noindent{\textbf{Explanation of Algorithmic Strategy:}}
The \textbf{XAI Snapshot view} (\cref{fig:xai_snapshot}) explains individual choices by comparing posterior means ($\mu$) and posterior draws ($\hat{\theta}$) across arms. The selected arm is always the one with the highest draw, even if its mean is not. At $t=228$, for example, Arm~8 had the highest mean, but Arm~7 was selected due to a higher draw. This led to a successful outcome (\cref{fig:teaser}C), updating its $\alpha$ (\cref{fig:teaser}B), sharply increasing belief (\cref{fig:teaser}A), and Arm~7 continues to be selected more frequently thereafter~(\cref{fig:teaser}C).

\noindent{\textbf{Understanding Sample Uncertainty and Risk:}}
The HDR Evolution Plot (\cref{fig:teaser}A) visualizes each arm’s posterior distribution over time. Posterior draws outside the central HDR reflect low-probability outcomes, signaling high epistemic uncertainty. TS-Insight reveals whether a choice reflects strong belief or a rare draw from a broad or unstable posterior.

\section{Conclusion}
TS-Insight offers a visual framework to reveal the internal mechanisms of TS and DTS. By assembling multiple plots and views, it supports the verification, diagnosis, and explanation, supporting trust in and more effective application of this family of algorithms. Future work includes online (real-time) visualization and formal user studies to quantify the tool's impact on debugging. We also plan to evolve the design towards fully interactive Coordinated Multiple Views, facilitating comparisons across arms.






\appendix
\section{Additional References}


\section{Limitations and Scope}
TS-Insight is explicitly designed for Thompson Sampling and its variants (e.g., Discounted TS, Sliding-Window TS, and Batched TS). The tool is not intended for contextual, combinatorial, or non-Bayesian bandits (e.g., UCB, EXP3), and its visualizations do not generalize to these paradigms.
\section{Key Concepts Explained}
\label{appendix:concepts}

\noindent This section provides simple, grounded explanations for some of the core technical terms used in this paper.

\paragraph{\textbf{The Multi-Armed Bandit Problem.}}
\label{appendix:mab-rl}

\noindent The Multi-Armed Bandit (MAB) problem is a canonical framework in sequential decision-making under uncertainty. Originally framed as a metaphor for a gambler facing a row of slot machines (``one-armed bandits''), each with an unknown probability of reward, the agent's goal is to maximize cumulative gain over time by learning which arm to pull.

\noindent Formally, at each time step $t \in \{1, 2, \dots\}$, the agent selects an action (arm) $a_t \in \mathcal{A}$ and observes a stochastic reward $r_t \sim \mathbb{P}(r \mid a_t)$, where $\mathcal{A}$ denotes the set of $K$ available arms.
This distilled setting captures the fundamental \textbf{exploration--exploitation dilemma}:
\begin{itemize}
    \item \textbf{Exploitation:} Favoring actions believed to yield the highest expected reward, leveraging current estimates.
    \item \textbf{Exploration:} Selecting actions with greater uncertainty to refine knowledge, potentially uncovering better long-term strategies.
\end{itemize}

\noindent MABs are powerful abstractions behind several real-world decision systems, including recommender systems, clinical trials, adaptive routing, and active learning.

\paragraph{\textbf{Thompson Sampling (TS).}}
Thompson Sampling is a sophisticated strategy for solving the MAB problem. Instead of just keeping a simple average of wins for each machine, it maintains a full probability distribution, a "belief", about how good each machine might be. To make a choice, it doesn't just pick the machine with the highest average; it takes a random sample from its belief about every machine and pulls the lever of the one with the highest sample. This naturally balances exploration and exploitation: machines with high uncertainty will sometimes produce very high "lucky" samples, causing the algorithm to explore them.

\paragraph{\textbf{Posterior Distribution (or "Belief").}}
In Bayesian terms, this is the updated belief about an arm's success rate after observing some data (i.e., after pulling the lever a few times). In our case, it's a Beta distribution, which is well-suited for modeling the probability of a binary outcome (like success/failure). It is defined by two numbers, $\alpha$ (related to successes) and $\beta$ (related to failures).

\paragraph{\textbf{Posterior Draw.}}
This is a single random number sampled from the posterior distribution. It represents one plausible guess for what the arm's true success rate might be, given the current evidence. Thompson Sampling's core mechanic is to make one such "guess" for every arm and then act on the most optimistic guess.

\paragraph{\textbf{Highest Density Region (HDR).}}
The HDR is a way to summarize a probability distribution. A 50\% HDR for an arm's belief is the range of values where we are 50\% certain the arm's true success rate lies. Visually, it's the "green" part of the distribution. In our tool, it represents the "high-confidence" or "typical" region of an arm's belief.

\section{Computing the Highest Density Region (HDR) for a Beta Posterior}

Let $\theta\sim\mathrm{Beta}(\alpha,\beta)$ be the posterior over an arm’s true success probability, where $\alpha>0$, $\beta>0$.  We wish to compute a symmetric $\rho$‐level Highest Density Region (HDR), i.e.\ the interval
\[
	\mathrm{HDR}_{\rho}(\alpha,\beta)\;=\;\bigl[a_{\rho}\,,\,b_{\rho}\bigr]
\]
such that
\[
	P\bigl(\theta\in[a_{\rho},\,b_{\rho}]\bigr)
	\;=\;\int_{a_{\rho}}^{b_{\rho}}
	f_{\mathrm{Beta}}(x;\,\alpha,\beta)\,dx
	\;=\;\rho,
\]
and for every $x\in[a_{\rho},\,b_{\rho}]$ and $y\notin[a_{\rho},\,b_{\rho}]$,
\[
	f_{\mathrm{Beta}}(x;\,\alpha,\beta)\;\ge\;
	f_{\mathrm{Beta}}(y;\,\alpha,\beta).
\]
In practice, we approximate a symmetric HDR around the posterior mean
\[
	\mu \;=\;\frac{\alpha}{\,\alpha+\beta\,}\,,
\]
by finding the smallest $\delta\ge0$ satisfying
\[
	F_{\mathrm{Beta}}\bigl(\mu+\delta\,;\,\alpha,\beta\bigr)
	\;-\;
	F_{\mathrm{Beta}}\bigl(\mu-\delta\,;\,\alpha,\beta\bigr)
	\;=\;\rho,
\]
where $F_{\mathrm{Beta}}(\cdot;\,\alpha,\beta)$ denotes the CDF of the $\mathrm{Beta}(\alpha,\beta)$ distribution.  Equivalently:
\begin{equation}
	\label{eq:hdr-def}
	a_{\rho} \;=\;\max\bigl\{\,0,\;\mu - \delta^*\bigr\}, 
	\qquad
	b_{\rho} \;=\;\min\bigl\{\,1,\;\mu + \delta^*\bigr\},
\end{equation}
with
\[
	\delta^* 
	\;=\;
	\arg\min_{\delta\ge0}\;
	\Bigl\{\, 
	F_{\mathrm{Beta}}\bigl(\mu+\delta;\,\alpha,\beta\bigr)
	\;-\;
	F_{\mathrm{Beta}}\bigl(\mu-\delta;\,\alpha,\beta\bigr)
	\;-\;\rho
	\,\Bigr\}^{2}.
\]
In our implementation, we solve for $\delta^*$ by a simple bisection loop:

\begin{enumerate}
  \item Initialize
    $\displaystyle \delta_{\min} = 0$, \quad 
    $\displaystyle \delta_{\max} = \min\{\mu,\,1-\mu\}$,
    \quad and tolerance $\varepsilon>0$.

  \item While $\delta_{\max} - \delta_{\min} > \varepsilon$:
    \begin{enumerate}
      \item Let $\delta = \tfrac12\,(\delta_{\min} + \delta_{\max})$.

      \item Compute 
      \[
        L = F_{\mathrm{Beta}}(\mu - \delta;\,\alpha,\beta), 
        \quad
        U = F_{\mathrm{Beta}}(\mu + \delta;\,\alpha,\beta).
      \]
      \item If $(U - L) \;>\;\rho$, then set $\delta_{\max} = \delta$; 
            else set $\delta_{\min} = \delta$.
    \end{enumerate}

  \item Return $a_{\rho} = \max\{0,\,\mu - \delta_{\max}\}$ and 
               $b_{\rho} = \min\{1,\,\mu + \delta_{\max}\}$.
\end{enumerate}

\noindent For the special cases where $\alpha \le 0$ or $\beta \le 0$ (which occur only
transiently when, e.g., a pure prior has no effective observations),
or when $\mu\in\{0,1\}$ so that $\delta_{\max}=0$, we simply set
\[
	a_{\rho} \;=\; b_{\rho} \;=\; \mu.
\]
Hence, at each time step $t$, given $(\alpha_t,\beta_t)$ for arm $k$, we compute
\[
	\bigl[a_{t},\,b_{t}\bigr] 
	\;=\;\mathrm{HDR}_{\rho}\bigl(\alpha_t,\;\beta_t\bigr)
\]
and shade the “central” band between $a_{t}$ and $b_{t}$ (often in green),
while coloring the two “tails” $(0,a_t)$ and $(b_t,1)$ in gray.  The resulting
50 \% HDR visually conveys the most probable range of $\theta$ for arm $k$ at step $t$.
\end{document}